\pdfoutput=1

\documentclass[11pt]{article}

\usepackage[final]{acl}

\usepackage{times}
\usepackage{latexsym}

\usepackage[T1]{fontenc}

\usepackage[utf8]{inputenc}

\usepackage{microtype}

\usepackage{inconsolata}

\usepackage{graphicx}
\usepackage{amsmath}
\usepackage{amssymb}
\usepackage{booktabs}
\usepackage{algorithmic}
\usepackage{amsfonts}
\usepackage{subfigure}
\usepackage{url}
\usepackage{multirow}
\usepackage{multicol}
\usepackage{arydshln}
\usepackage{bbm}
\usepackage[ruled,vlined]{algorithm2e}
\usepackage{colortbl}
\usepackage[most]{tcolorbox}
\tcbuselibrary{skins, breakable}

\usepackage{dsfont}

\title{LearnLens: LLM-Enabled Personalised, Curriculum-Grounded Feedback with Educators in the Loop}

\author{Runcong Zhao$^{1}$, Artem Bobrov$^{1}$,  Jiazheng Li$^1$, Cesare Aloisi$^2$, Yulan He$^{1}$\\
  $^1$King's College London,  $^2$AQA
  \\
  \texttt{\{runcong.zhao, artem.bobrov, jiazheng.li\}@kcl.ac.uk} \\ \texttt{caloisi@aqa.org.uk, yulan.he@kcl.ac.uk} } 

\begin{document}
\maketitle
\def\thefootnote{\arabic{footnote}}
\begin{abstract}

Effective feedback is essential for student learning but is time-intensive for teachers. We present \textbf{LearnLens}, a modular, LLM-based system that generates personalised, curriculum-aligned feedback in science education. LearnLens comprises three components: (1) an \emph{error-aware assessment} module that captures nuanced reasoning errors; (2) a \emph{curriculum-grounded generation} module that uses a structured, topic-linked memory chain rather than traditional similarity-based retrieval, improving relevance and reducing noise; and (3) an \emph{educator-in-the-loop} interface for customisation and oversight. LearnLens addresses key challenges in existing systems, offering scalable, high-quality feedback that empowers both teachers and students. A screencast video introducing the system\footnote{ \url{https://www.youtube.com/watch?v=mCUALVwDKNQ}} and the demo\footnote{\url{https://learnlens.co.uk/login} 
} are available online.

\end{abstract}

\section{Introduction}

Timely, high-quality feedback is a key driver of learning across all educational levels \citep{hattie2007power, boud2023feedback}. Effective feedback should be personalised, actionable, and dialogic: supporting student engagement and offering clear guidance for improvement \citep{carless2016dialogue}. However, delivering such feedback at scale remains a major challenge, as it is time-intensive and adds to teachers’ already substantial workload and growing levels of stress and fatigue \citep{jomuad2021teachers}.
Recent advances in large language models (LLMs) offer a promising opportunity to automate feedback generation and ease teacher workload. Prior work \citep{mazzullo2025automated, liu2025survey} has explored their use for automatically generating scores and rationales for student answers, as well as for delivering evidence-based feedback in classroom settings. 
However, existing systems still fall short in several ways: (i) they focus narrowly on scoring, overlooking the learner’s partial understanding and reasoning process \citep{mayfield2020fine, xie2022automated}; (ii) they generate feedback directly from LLMs without sufficient curricular grounding, often leading to hallucinated, misaligned, or overly generic suggestions \citep{mazzullo2025automated, meyer2024using}; and (iii) they offer limited avenues for educator control, correction, or customisation during the feedback process \citep{swamy2025illuminate}.

  \begin{figure*}[t!]
    \centering
    \includegraphics[width=\linewidth]{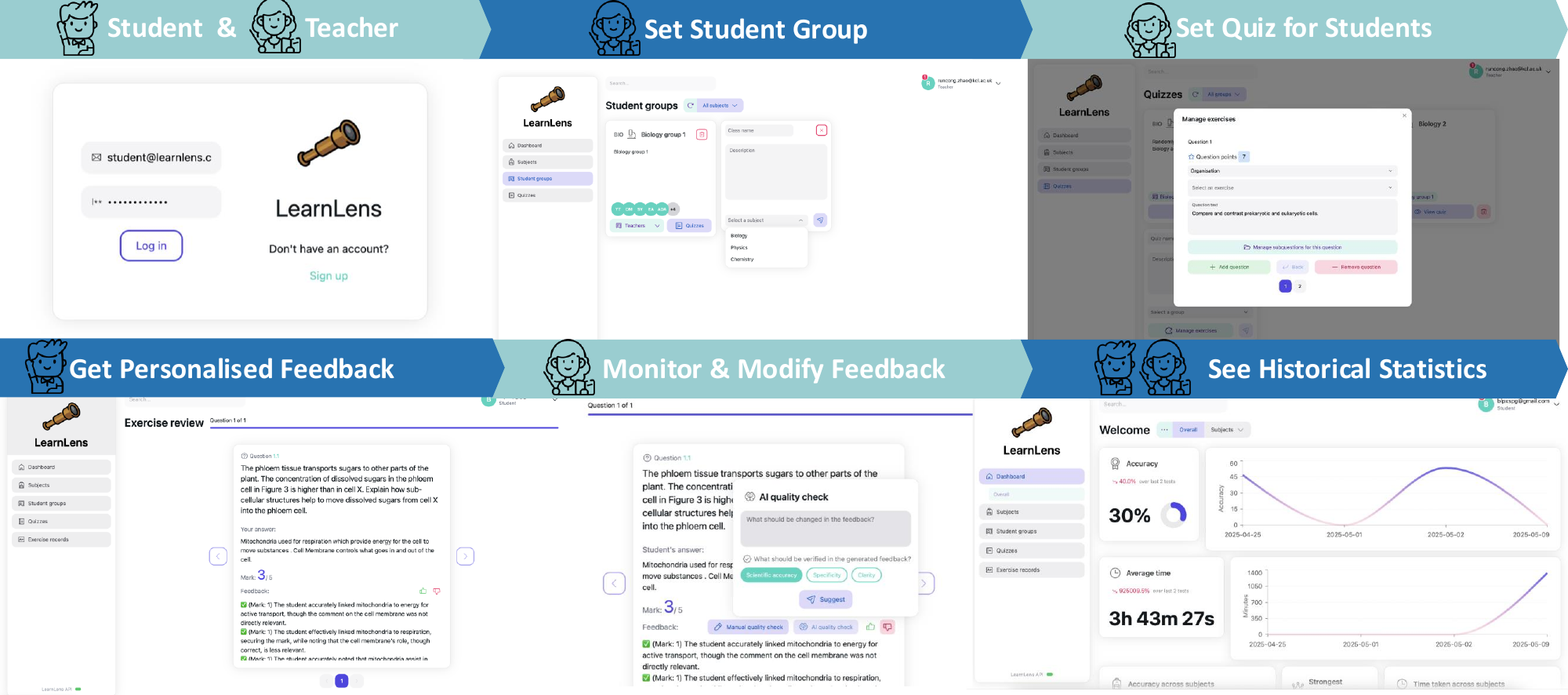}
    \caption{\small {\textbf{LearnLens}: A modular LLM-based system delivering personalised, curriculum-aligned feedback through error analysis, topic memory, and educator oversight.}}
    \label{fig:learnlens}
\end{figure*}

To address these limitations, we present \textbf{LearnLens}, a modular LLM-based system for personalised, curriculum-aligned feedback in science education. LearnLens consists of three core components, each directly tackling a corresponding shortcoming. First, an \emph{error-aware assessment} module moves beyond binary correctness by aligning learner responses with structured mark schemes and identifying conceptual, factual, or linguistic errors, enabling more nuanced understanding of student reasoning. Second, a \emph{curriculum-grounded generation} module produces feedback using a restructured, topic-linked memory chain aligned with the national curriculum. Unlike traditional similarity-based retrieval, which often introduces irrelevant or noisy content, our \textit{Chain-of-Concept} framework integrates carefully curated knowledge and curriculum alignment into the retrieval process, ensuring that generated feedback is pedagogically coherent and tailored to specific learning objectives. Third, an \emph{educator-in-the-loop} interface, paired with a separate student-facing interface, enables teachers and students to interact with the system in complementary ways. Teachers can monitor student performance, create customised questions, review and revise feedback using natural language, and optionally select from a suite of embedded verifiers that assess different aspects of feedback quality—such as factual accuracy, curricular relevance, and linguistic clarity.

 \section{Architecture of LearnLens}

We design \textbf{LearnLens} as a dual-interface intelligent feedback system that serves both teachers and students. Our core goal is twofold: (1) to reduce the workload for teachers while enhancing their ability to monitor student progress, and (2) to provide students with timely, personalised feedback that supports deeper learning. Powered by LLMs, \textbf{LearnLens} integrates structured knowledge management and human-in-the-loop collaboration into a scalable educational tool. Through dedicated teacher and student interfaces, the system delivers curriculum-aligned, high-quality support tailored to real classroom needs.

\begin{figure*}[h]
    \centering
    \includegraphics[width=0.9\linewidth]{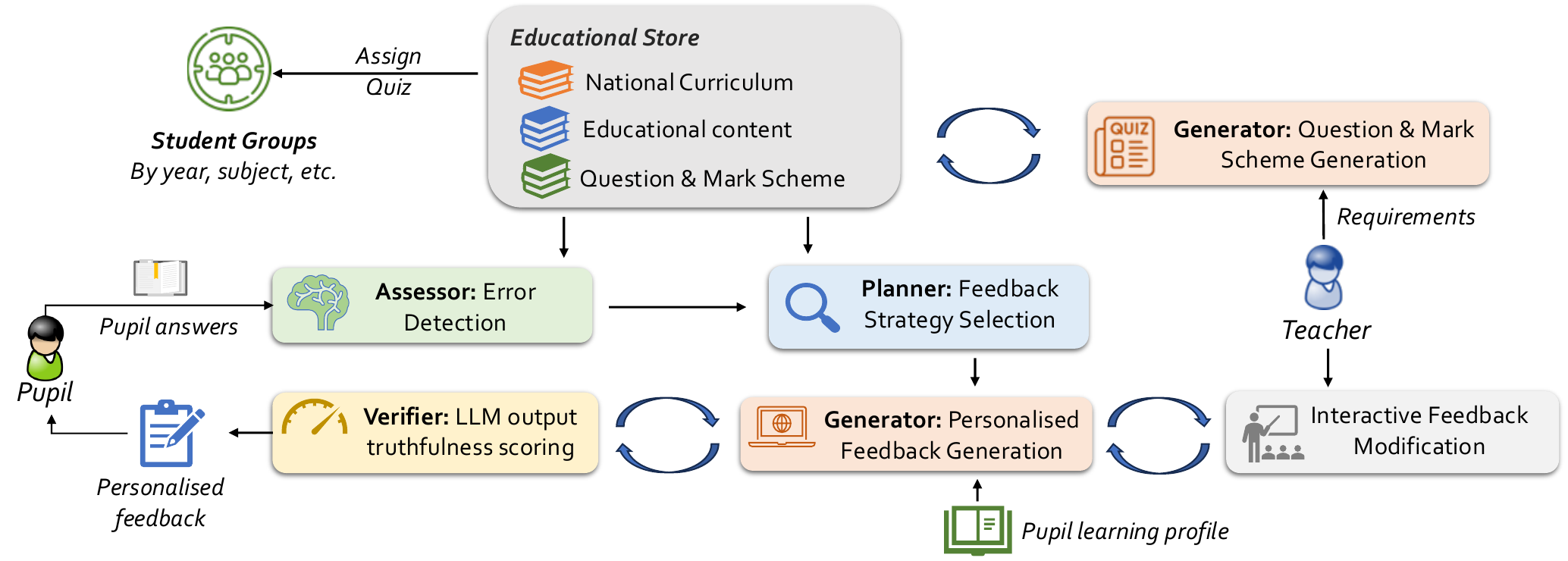}
    \caption{\texttt{LearnLens}: Overall framework.}
    \label{fig:toolFramework}
\end{figure*}

\subsection{Teacher Interface: Less Grading, More Guiding}
To reduce teacher workload while enhancing instructional oversight, \textbf{LearnLens} provides tools that support differentiated instruction, personalised assessment, and high-quality feedback. As shown in Figure~\ref{fig:learnlens}, the teacher interface facilitates efficient classroom management through student grouping, quiz creation, and curriculum-aligned rubric generation. It also offers up-to-date analytic insights, including performance visualisation and verifier scores, to help teachers identify students in need of support and feedback requiring revision. This integrated, human-in-the-loop workflow enables teachers to focus less on repetitive grading and more on delivering targeted, meaningful guidance.

\paragraph{Student Grouping} 
As illustrated by the \emph{``Set Student Group''} module, teachers can organise students into groups based on subject, year level, and learning progress. Each group can then be assigned one or more teachers and a cohort of students, with appropriate permissions configured to control access to shared materials and feedback within that group. This structure enables differentiated instruction, collaborative planning, and role-based access control in a scalable and intuitive way.

\paragraph{Quiz Creation}
As illustrated by the \emph{``Set Quiz for Students''} module, teachers can create quizzes and assign them to specific student groups. Each question is first associated with a topic from the national curriculum, ensuring alignment with prescribed learning objectives. For each selected topic, the system provides a curated bank of pre-authored questions that teachers can readily select. Alternatively, teachers may compose their own custom questions to address specific instructional goals. To further enhance flexibility and reduce authoring effort, teachers can also prompt the LLM to automatically generate questions based on the chosen topic and specified pedagogical requirements.

\paragraph{Curriculum-Aligned Rubrics}
To ensure consistency in model assessment, the system first generates a mark scheme for each question. Since the mark scheme impacts all subsequent assessments, we consider it a critical step where accuracy is prioritised over time efficiency. In line with inference-time scaling laws \citep{snell2024scalingllmtesttimecompute, brown2024largelanguagemonkeysscaling}, the model is granted extended reasoning time to explore diverse solution paths and consider multiple correct approaches. To further ensure pedagogical validity, the system retrieves relevant curriculum content for the target topic. This enables the model to distinguish between knowledge that students are expected to have mastered and content beyond their current level, thereby constraining the mark scheme within the appropriate learning scope.

\paragraph{Student Performance Visualisation}
To support effective teacher intervention, \textbf{LearnLens} provides a historical performance visualisation module that highlights students who may require additional support. Prior studies~\citep{OzerOzkan2025Decoding} and our interviews with teachers suggest that educators often struggle to monitor students’ mastery levels in a timely manner. This module addresses that challenge by offering a clear, data-driven overview of student progress, enabling teachers to more efficiently identify learning gaps.

\paragraph{Verifier for Quality Assessment}
To track the quality of model-generated feedback and assist teachers in identifying responses that may require attention, each feedback instance is first self-evaluated against a set of educational criteria: \textit{Scientific Accuracy} (whether misconceptions are correctly identified and explanations are valid), \textit{Clarity} (whether the language is accessible to GCSE-level students\footnote{LearnLens can be tailored to students across all educational levels. In this demo, we focus on GCSE science.}), and \textit{Specificity} (whether the feedback clearly highlights what is correct, incorrect, or missing). For each criterion, the model assigns a self-verification score along with a justification. 

As shown in Figure~\ref{fig:toolFramework}, \textbf{LearnLens} integrates a verification-and-revision loop into its assessment pipeline. To mitigate model-family bias, verification is performed by a separate foundation model from the generator. Feedback is iteratively refined based on verifier scores: if a response falls short, the model identifies weaknesses and revises it. The process continues until all criteria meet a threshold or the iteration limit is reached:
\[
\text{stop} \;=\; \bigl[\min_i r_i \ge \tau\bigr]
                 \;\lor\; \bigl[t \ge T_{\text{max}}\bigr],
\]
where \(r_i\) is the verifier score for the \(i\)-th criterion, \(\tau\) is the acceptance threshold, and \(t\) is the current iteration. The highest-scoring feedback is returned. Examples with verification scores are shown in Appendix~\ref{sec:personalised-feedbacks}.

\paragraph{Interactive Feedback Modification}
\textbf{LearnLens} integrates student performance visualisation and verifier scores to help teachers efficiently identify feedback that may require revision. 
Teachers can refine flagged feedback through a conversational interface, either by directly editing the content or by providing high-level suggestions. These modifications can be applied to a single question or propagated across the entire quiz. The AI agent then incorporates the teacher’s input and generates an updated response accordingly. An illustrative example is provided in Appendix~\ref{sec:teacher-modification}.

Teacher intervention serves as a signal of dissatisfaction with the initial output. In such cases, the system dynamically allocates additional computational resources, allowing the model to reflect more deeply and produce higher-quality feedback. While faster refinement strategies were explored, they often compromised fidelity to teacher intent. Given this trade-off, we prioritise feedback quality to preserve user trust and learning effectiveness.

\subsection{Student Interface: Faster Feedback, Better Understanding}
From the student perspective, \textbf{LearnLens} delivers timely and tailored feedback that promotes continuous learning. To ensure the feedback is both accurate and actionable for learning, the system combines reliable assessment, memory-driven strategy planning, and safe, context-aware generation.

\paragraph{Assessor}

To deliver \emph{consistent, fine-grained} scoring, the Assessor maps each student
answer \(\hat{a}_i\) onto the curriculum-aligned mark scheme
\(\mathcal{C}_i=\{(c_k,w_k)\}_{k=1}^{K_i}\),
where \(c_k\) is the \(k\)-th key concept and \(w_k\in \mathbb{Z}_{>0}\) its weight.
The raw score is the weighted sum of concept matches:\[
s_i \;=\; \sum_{k=1}^{K_i} w_k\;
      \textsc{Match}(c_k,\hat{a}_i),\]
with \textsc{Match} satisfied when the LLM detects that the concept
appears in the answer, thus granting partial credit even if the final
answer is wrong.  
For internal consistency we also obtain a \emph{prompt score}
\(s_i^{\text{prompt}}\) via a direct “grade-this-answer’’ prompt and
trigger self-reflection whenever \(s_i \neq s_i^{\text{prompt}}\) \citep{li-etal-2024-calibrating}.
LearnLens is also robust to surface
errors: grammatical or typographical mistakes do \emph{not} affect the
score \(s_i\). Such issues are instead captured by a separate
expression-quality flag \(\delta_i\in\{0,1\}\) (1 = major language
issues), which is shown in the feedback but excluded from the numerical
grade. 
This design (i) renders the weighting and partial credit scheme \emph{transparent}: students and teachers can directly inspect the marks \(w_k\) allotted to each concept \(c_k\) and see how many were awarded, which removes the black box impression and eases disputes; and (ii) decouples conceptual understanding from writing quality.

\begin{figure}[h]
    \centering
    \includegraphics[width=\linewidth]{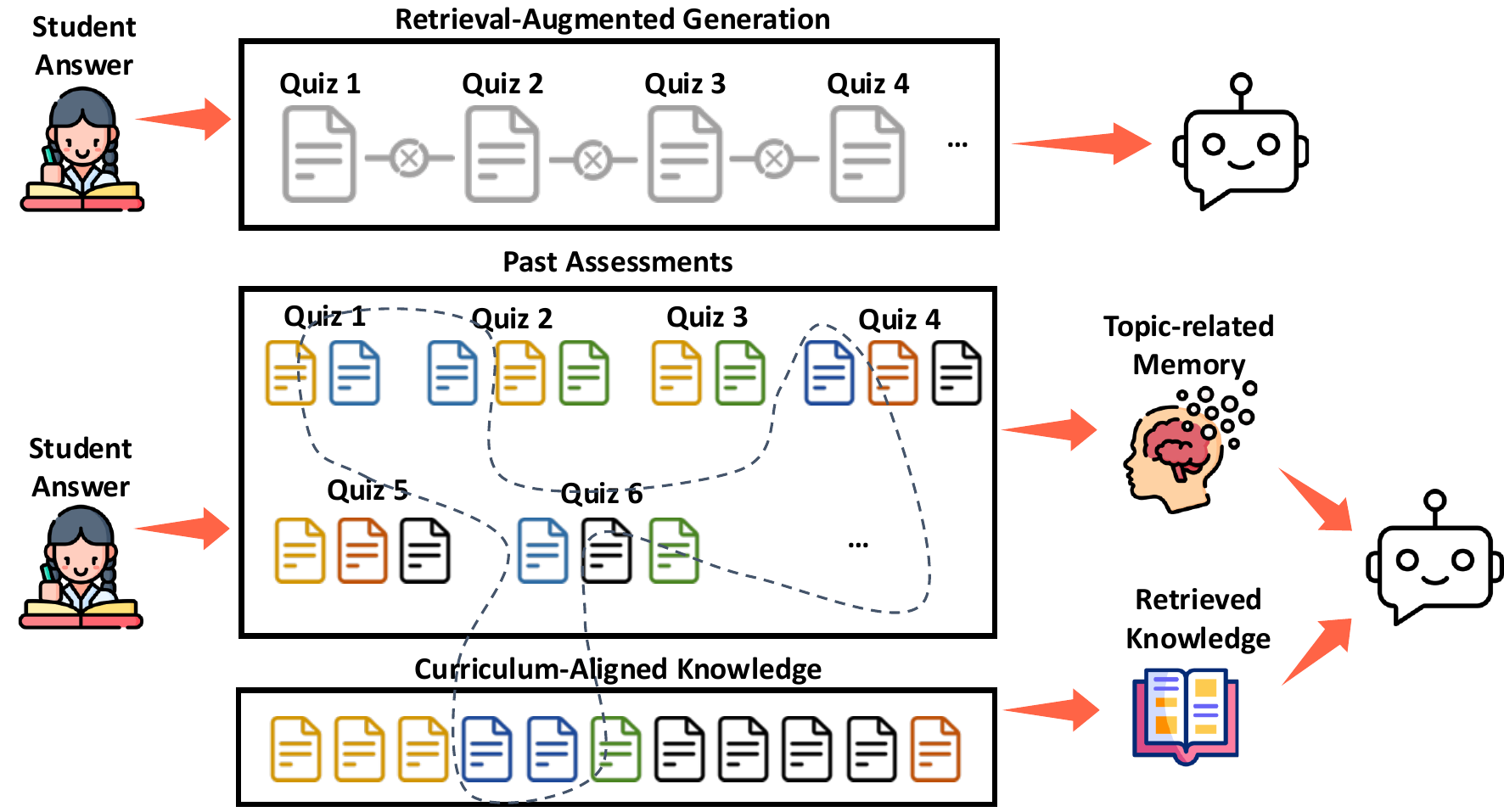}
    \caption{\small {Comparison between traditional RAG and the proposed Chain-of-Concept approach. In RAG (top), past assessment records are stored sequentially without explicit logical structure, forcing the retriever to rely solely on surface-level similarity. This often introduces significant noise, especially as the database grows. In contrast, our approach (bottom) organises past assessments by topic-level relationships, enabling the model to retrieve more contextually relevant information and generate more personalised, coherent feedback.}}
    \label{fig:memory}
\end{figure}

\paragraph{Planner} 
As shown in Figure~\ref{fig:memory}, we organise past assessments by \emph{curriculum topics} rather than raw embedding similarity used in earlier work \citep{gao2023retrieval}. Each question is decomposed into sub-questions (nodes) \(v_i\) and labelled with one or more topics \(\mathcal{L}(v_i)\subseteq\mathcal{T}\). We model these records as a topic graph 
\begin{equation*}
\resizebox{0.95\linewidth}{!}{$
  G = (V,E,\mathcal{T}),\quad
  E = \{(v_i,v_j)\mid \mathcal{L}(v_i)\cap\mathcal{L}(v_j)\neq\varnothing\}.
$}
\end{equation*}
For a query \(q\) tagged with topics \(C_q\subseteq\mathcal{T}\),
retrieval is confined to the topic subgraph 
\(
G_q=G\bigl[\{\,v_j \mid \mathcal{L}(v_j)\cap C_q\neq\varnothing\}\bigr],
\)
and we return
\[
\mathcal{R}(q)=\operatorname*{arg\,top\text{-}k}_{v_j\in G_q} f(q,v_j),
\]
where \(f\) is a FAISS-based ranker \citep{douze2024faiss}. 
Topic filtering removes cross-topic noise, yielding more coherent evidence; the module then analyses pupil errors and prior knowledge to choose the most effective feedback strategy.

\paragraph{Generator} 
To generate high-quality feedback, \textbf{LearnLens} integrates the quiz content, the corresponding mark scheme, and the selected feedback strategy. We adopt an LLM-based self-reflection mechanism~\citep{asai2024selfrag, zhang-etal-2024-self-contrast}, allowing the model to reason over the student's response. To ensure the safety and appropriateness of the generated content, a safety-aligned model is applied post-generation to filter out any harmful, biased, or otherwise inappropriate language.

\section{Evaluation}

\subsection{Experimental Setup}
We conducted a comparative evaluation of LLMs across multiple model families and sizes. Due to student data privacy concerns, all experiments were conducted via local deployment. Our evaluation includes Meta-Llama-3-8B-Instruct, Qwen2.5-32B-Instruct, and QwQ-32B.

\subsection{Evaluation Results}

 \begin{figure*}[ht!]
    \centering
    \includegraphics[width=\linewidth]{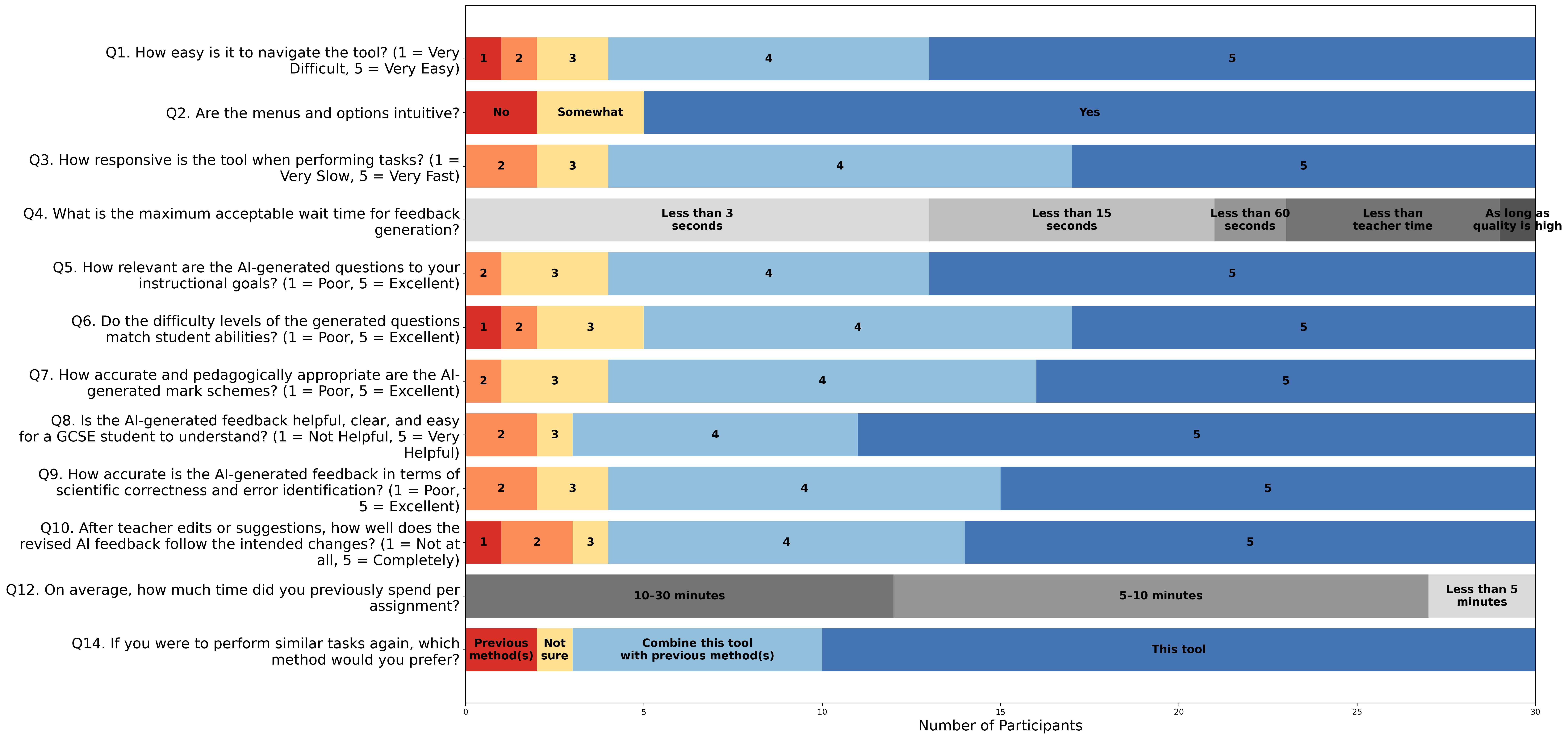}
    \caption{\small{Full Questionnaire Response Distribution (N = 30 per Question). Shades of blue denote more favourable evaluations of the tool, while shades of red indicate dissatisfaction. Grey segments represent neutral factual responses that are not direct indicators of user sentiment toward the tool.}}
    \label{fig:questionnaire}
\end{figure*}

\subsubsection{User Feedback Results}
To evaluate how well \textbf{LearnLens} aligns with real-world teaching needs, we collected responses from \emph{30 participants}, all of whom had prior teaching experience in a STEM module at either high school or college level. Each participant interacted with the system and completed a structured questionnaire (Figure~\ref{fig:questionnaire}), which covered topics such as interface usability, system responsiveness, question quality, feedback usefulness, and overall user experience. The complete questionnaire is provided in Appendix~\ref{sec:questionnaire}.
 \begin{figure*}[ht!]
    \centering
    \includegraphics[width=0.85\linewidth]{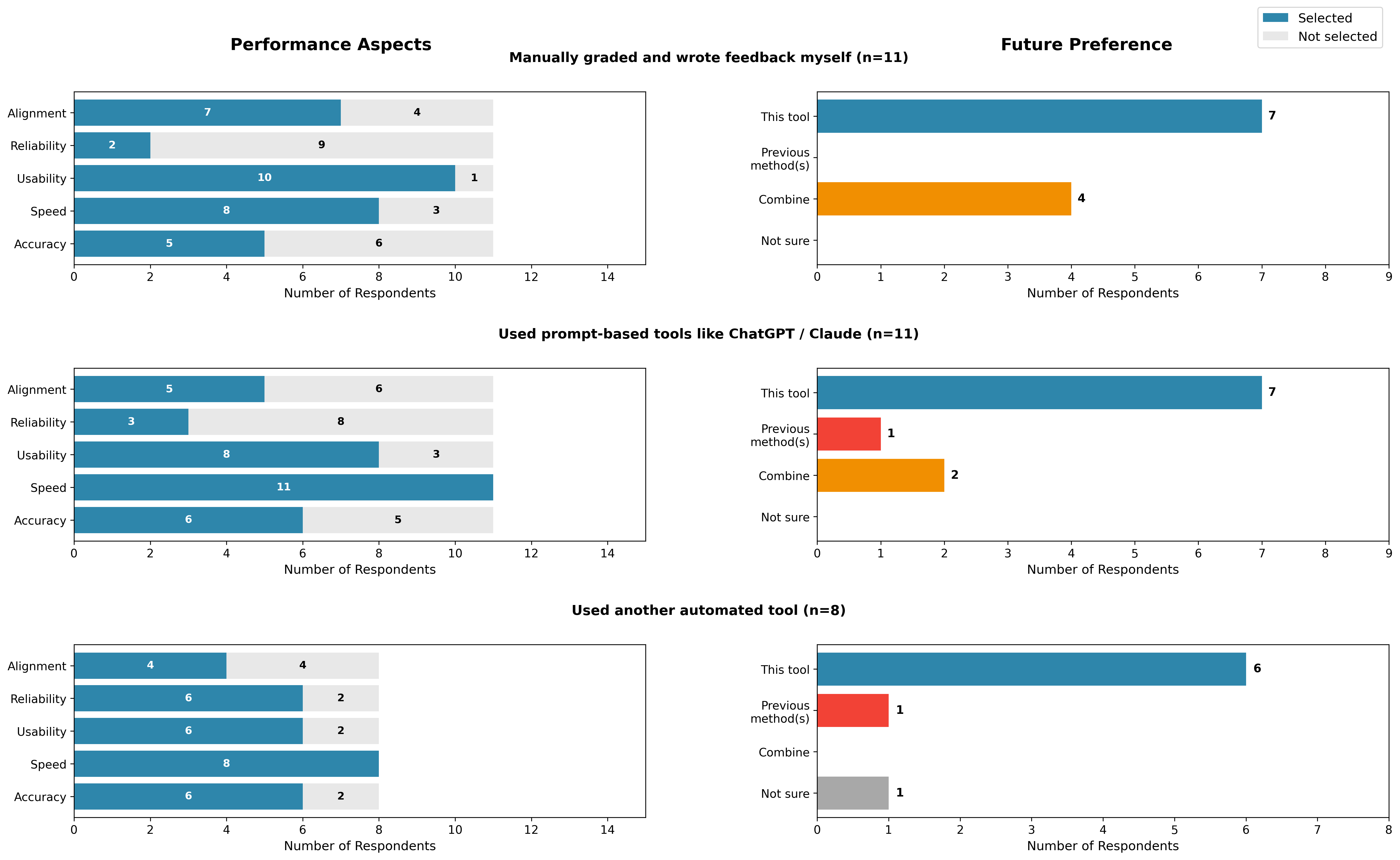}
    \caption{\small{Grouped feedback by participants’ prior methods (manual, prompt-based, or automated). Left: perceived improvements across five dimensions. Right: future preferences for task completion. Blue = positive toward our tool; red = preference for previous methods; orange = combine both; grey = neutral or other.}}
    \label{fig:grouped_feedback_comparison}
\end{figure*}
Our questionnaire results, summarised in Figure~\ref{fig:questionnaire} (full response distribution) and Figure~\ref{fig:grouped_feedback_comparison} (grouped comparison), highlight the following themes: 
\begin{table*}[t!]
\centering
\resizebox{0.7\linewidth}{!}{%
\begin{tabular}{lcccccccc}
\toprule
& \multicolumn{4}{c}{\textbf{Performance}} 
& \multicolumn{4}{c}{\textbf{Generation Latency \& Cost}} \\
\cmidrule(lr){2-5}\cmidrule(lr){6-9}
\textbf{Model} & \textbf{MSE} & \textbf{Corr.} & \textbf{Acc.} & \textbf{$\pm$1 Acc.} & 
\textbf{Avg.\,(s)} & \textbf{Min.\,(s)} & \textbf{Max.\,(s)} & \textbf{Cost\,(\$)} \\
\midrule
LLaMA-3-8B-Ins   & 3.468 & 0.235 & 0.190 & 0.646 &  5.14 &  2.43 &  9.40 & 0.0040 \\
Qwen2.5-32B-Ins  & 3.658 & 0.230 & 0.241 & 0.633 & 12.39 &  5.78 & 20.29 & 0.0096 \\
QwQ-32B          & 3.506 & 0.323 & 0.329 & 0.709 & 45.49 & 32.27 & 57.36 & 0.0351 \\
LearnLens             & \textbf{3.190} & \textbf{0.388} & \textbf{0.354} & \textbf{0.747} & 11.39 &  7.59 & 16.97 & 0.0099 \\
\bottomrule
\end{tabular}}
\caption{\textbf{LearnLens} performance and efficiency. MSE: mean squared error; Corr.: Pearson correlation; Acc.: exact match; $\pm1$ Acc.: within–one–mark accuracy. Avg./Min./Max.\ are generation latencies; Cost is per request.}
\label{tab:combined_metrics}
\end{table*}
\paragraph{Global Satisfaction and Usability}
Across the nine Likert-scale items (Q1, Q3, Q5–Q10) the mean rating never falls below 4.1/5, signalling a uniformly positive reception. Navigation ease and perceived responsiveness confirm that the dual-interface design achieves its principal UX goal: immediate, low-friction access to feedback tools. Likewise, high scores for relevance to instructional goals (4.4/5) and scientific accuracy (4.3/5) validate our curriculum-grounding strategy.

Categorical items echo this trend. About 80\% of teachers describe the menu structure as ``intuitive'', and 75\% expect results in under 15 seconds, a threshold LearnLens already meets in >90 \% of cases during local deployment. These data confirm that the system’s latency, a common pain-point for classroom AI, is acceptable for real-world use.

\paragraph{Efficiency Gains}
Q12 shows that half of teachers spent 10–30 minutes per assignment, whereas the median with \textbf{LearnLens} falls below 5 minutes. Q13 echoes this: 26/30 respondents rank \emph{speed} as the top strength. By off-loading rubric matching and error spotting, \textbf{LearnLens} shifts teachers’ time from grading to guidance, realigning workload rather than merely automating it.

\paragraph{Experience-Sensitive Priorities}
Figure \ref{fig:grouped_feedback_comparison} segments perceptions by prior practice. Manual graders value speed (91 \%) and usability (82 \%) most, unsurprising given their baseline workflow. Prompt-tool users distribute credit more evenly, highlighting LearnLens’ balanced improvement across accuracy, reliability, and alignment. Their familiarity with LLM quirks likely sharpens expectations. Other-automation users rank accuracy (88 \%) and reliability (75 \%) highest, suggesting that our verifier-in-the-loop architecture successfully addresses limitations they encountered elsewhere.

A $\tilde{\chi}^2$ test ($\alpha$ = 0.05) shows no significant difference in overall adoption intent across groups (p = 0.63); every cohort exceeds 65 \% ``continue using'', with the automation-savvy group peaking at 85 \%. Thus, while priorities diverge, and willingness to integrate LearnLens is stable.

\paragraph{Implications for Deployment} To translate these findings into actionable next steps, we highlight three deployment priorities that will maximise user satisfaction and long-term adoption:
(1) Latency ceiling. Meeting the sub-15-second expectation is critical; ongoing optimisation should therefore focus on inference batching and on-device caching.
(2) Adaptive onboarding. Manual graders respond to time-saving narratives, whereas technical users are swayed by demonstrable accuracy safeguards; onboarding materials should branch accordingly.
(3) Verifier transparency. High appreciation for accuracy among automation veterans highlights the value of surfacing verifier scores (§2.1); exposing these metrics to all users could further bolster trust.

In summary, the study confirms that LearnLens delivers broadly perceived pedagogical value, but supporting diverse existing workflows is crucial for its scalable adoption in classrooms.

\subsection{Agent Performance and Efficiency}
Table~\ref{tab:combined_metrics} reports results on \emph{100 authentic student answers} to a five-mark short-essay science question.  
Baselines use the same foundation models as those employed in LearnLens, but perform both feedback generation and verification through direct prompts to the same model, without modular decomposition. For fairness, all methods share the same inputs (question, mark scheme, answer) and required outputs (score, feedback, verification).

LearnLens follows a \emph{modular} pipeline: lightweight models handle routine subtasks, while larger models are invoked only when deeper reasoning is needed.  Combined with efficient \texttt{vLLM} serving and speculative decoding, this design \emph{achieves the strongest scoring quality without additional overhead}: MSE drops to 3.19 (8–13 \% below any baseline), correlation rises to 0.388, and both exact and $\pm1$ accuracies surpass the best baseline (QwQ-32B).  At the same time, average generation latency is 11.4 s, comparable to Qwen2.5-32B and four times faster than QwQ-32B, while the per-request cost (\$0.0099) matches Qwen2.5-32B and is 72 \% cheaper than QwQ-32B.

\section{Conclusion}
We introduce \textbf{LearnLens}, a dual-view feedback system that supports both teachers and students through curriculum-aligned assessment and personalised feedback. By combining LLM capabilities with human-in-the-loop design, LearnLens reduces teacher workload and enhances student learning, offering a practical path toward scalable, classroom-ready AI assistance.
While LearnLens shows promise in supporting teachers, we acknowledge the lack of student evaluation and plan to address this in future work.

\section*{Acknowledgments}
This work was supported by the UK Department for Education through the AI Catalyst Fund - AI Tools for Education (grant no. 10144187) and the UK Engineering and Physical Sciences Research Council (EPSRC) through a Turing AI Fellowship (grant no. EP/V020579/1, EP/V020579/2) and a Prosperity Partnership project with AQA (UKRI566). The authors also acknowledge the use of the King’s Computational Research, Engineering, and Technology Environment (CREATE) at King’s College London. 

\bibliography{custom}

\appendix

\section{Example: Personalised Feedback with Self-Verification Scores}
\label{sec:personalised-feedbacks}

To illustrate how \textbf{LearnLens} produces feedback with built-in quality assurance, we present an example of AI-generated personalised feedback accompanied by self-verification scores. The system evaluates student responses against a mark scheme, assigns partial credit where appropriate, and generates targeted feedback. Verification scores reflect the feedback’s scientific accuracy, clarity, and specificity, while reasoning statements explain how each score was determined.

\begin{tcolorbox}[title=Sample Feedback with Verification, colback=gray!5, colframe=gray!40!black, fonttitle=\small, fontupper=\small, breakable]

\textbf{AI-Generated Feedback:}
\begin{itemize}
    \item \textbf{Mark: 0} – Include that cell X contains mitochondria, which are needed to provide energy for active transport.
    \item \textbf{Mark: 1} – Correctly identified active transport moves sugars against the concentration gradient from low (cell X) to high (phloem).
    \item \textbf{Mark: 0} – Mention that mitochondria enable aerobic respiration to release energy specifically needed for active transport.
    \item \textbf{Mark: 1} – Correctly explained energy is needed to move sugars against the concentration gradient via active transport.
    \item \textbf{Mark: 1} – Correctly stated active transport moves sugars from low to high concentration in the phloem.
\end{itemize}

\vspace{0.5em}
\textbf{Total Mark:} 3/5

\vspace{1em}
\textbf{Verification Scores:}
\begin{itemize}
    \item \emph{Scientific Accuracy:} \textbf{5}
    \item \emph{Specificity:} \textbf{5}
    \item \emph{Clarity:} \textbf{4}
\end{itemize}

\vspace{0.5em}
\textbf{Verification Reasoning:}
\begin{itemize}
    \item \emph{Scientific Accuracy:} Full marks.
    \item \emph{Specificity:} Feedback addresses specific content of the student's answer by highlighting what was right, wrong, or missing.
    \item \emph{Clarity:} Language is generally clear and accessible to GCSE students, though some simplification could reduce potential confusion.
\end{itemize}

\end{tcolorbox}

\section{Example: Interactive Feedback Modification}
\label{sec:teacher-modification}

To illustrate how teacher input is incorporated into the feedback generation pipeline, we present an example of interactive feedback refinement. The original AI-generated feedback is revised based on a set of teacher-provided suggestions that emphasise clarity, brevity, and actionability. The differences introduced by the model in response to teacher suggestions are highlighted in red for clarity and ease of comparison.

\begin{tcolorbox}[title=Original AI Feedback, colback=gray!5, colframe=gray!40!black, fonttitle=\small, fontupper=\small, breakable]

\begin{itemize}
    \item \textbf{Mark: 0} – Add explicit mention of mitochondria in cell X. Specify that mitochondria in cell X facilitate aerobic respiration to provide energy \textcolor{red}{needed for active transport}.
    \item \textbf{Mark: 1} – Correctly identified active transport moving sugars against the concentration gradient from cell X to phloem.
    \item \textbf{Mark: 0} – Specify that energy released via respiration is required for active transport. Include \textcolor{red}{``to release energy needed for active transport.''}
    \item \textbf{Mark: 1} – Correctly explained movement against the concentration gradient via active transport. \textcolor{red}{Correct direction noted.}
    \item \textbf{Mark: 1} – Correctly stated active transport from low (cell X) to high concentration (phloem).
\end{itemize}

\end{tcolorbox}

\vspace{1em}

\textbf{Teacher Suggestions:}
\begin{itemize}
    \item Do not include general suggestions.
    \item Give short, specific instructions that students can implement.
    \item Keep the feedback concise.
\end{itemize}

\vspace{1em}

\begin{tcolorbox}[title=Revised Feedback (with Highlighted Edits), colback=gray!5, colframe=gray!40!black, fonttitle=\small, fontupper=\small, breakable]

\begin{itemize}
    \item \textbf{Mark: 0} – Specify that cell X contains many mitochondria, which are needed for aerobic respiration to produce energy for active transport. \textcolor{red}{(concise)}
    \item \textbf{Mark: 1} – Correctly identified active transport moving sugars against the concentration gradient from cell X to phloem. \textcolor{red}{(unchanged)}
    \item \textbf{Mark: 0} – Add that mitochondria carry out respiration to release energy specifically \textcolor{red}{for active transport of sugars}.
    \item \textbf{Mark: 1} – Correctly explained movement against the concentration gradient via active transport. \textcolor{red}{Direction (low~$\rightarrow$~high) is accurate.}
    \item \textbf{Mark: 1} – Correctly stated active transport from low (cell X) to high concentration (phloem). \textcolor{red}{(unchanged)}
\end{itemize}

\end{tcolorbox}

\section{User Questionnaire}
\label{sec:questionnaire}

To better understand user needs and evaluate the usability and effectiveness of \textbf{LearnLens}, we designed a structured questionnaire for teachers. The survey covered five key dimensions: interface design, system performance, question creation and assessment, personalised feedback quality, and overall user experience.  The full questionnaire is detailed below.

\vspace{1em}

\begin{tcolorbox}[title=Section 1: Interface Design, colback=gray!5, colframe=gray!40!black, breakable, fonttitle=\small, fontupper=\small]
\textbf{Q1.} How easy is it to navigate the tool? (1 = Very Difficult, 5 = Very Easy)

\vspace{0.8em}
\textbf{Q2.} Are the menus and options intuitive? \\
- Yes\\
- No \\
- Somewhat
\end{tcolorbox}

\begin{tcolorbox}[title=Section 2: Performance, colback=gray!5, colframe=gray!40!black, breakable, fonttitle=\small, fontupper=\small]
\textbf{Q3.} How responsive is the tool when performing tasks? (1 = Very Slow, 5 = Very Fast)

\vspace{0.8em}
\textbf{Q4.} What is the maximum acceptable wait time for feedback generation? \\
- Less than 3 seconds \\
- Less than 15 seconds \\
- Less than 60 seconds \\
- Less than the time a teacher would typically spend providing feedback \\
- As long as the feedback is high quality
\end{tcolorbox}

\begin{tcolorbox}[title=Section 3: Question Creation, colback=gray!5, colframe=gray!40!black, breakable, fonttitle=\small, fontupper=\small]
\textbf{Q5.} How relevant are the AI-generated questions to your instructional goals? (1 = Poor, 5 = Excellent)

\vspace{0.8em}
\textbf{Q6.} Do the difficulty levels of the generated questions match student abilities? (1 = Poor, 5 = Excellent)

\vspace{0.8em}
\textbf{Q7.} How accurate and pedagogically appropriate are the AI-generated mark schemes? (1 = Poor, 5 = Excellent)
\end{tcolorbox}

\begin{tcolorbox}[title=Section 4: Personalised Feedback, colback=gray!5, colframe=gray!40!black, breakable, fonttitle=\small, fontupper=\small]
\textbf{Q8.} Is the AI-generated feedback helpful, clear, and easy for a GCSE student to understand? (1 = Not Helpful, 5 = Very Helpful)

\vspace{0.8em}
\textbf{Q9.} How accurate is the AI-generated feedback in terms of scientific correctness and error identification? (1 = Poor, 5 = Excellent)

\vspace{0.8em}
\textbf{Q10.} After teacher edits or suggestions, how well does the revised AI feedback follow the intended changes? (1 = Not at all, 5 = Completely)

\end{tcolorbox}

\begin{tcolorbox}[title=Section 5: Overall Experience \& Suggestions, colback=gray!5, colframe=gray!40!black, breakable, fonttitle=\small, fontupper=\small]
\textbf{Q11.} Before using our tool, how did (or would) you complete similar tasks? (Select all that apply) \\
- Manually graded and wrote feedback myself \\
- Used prompt-based tools like ChatGPT / Claude \\
- Used another automated tool \\
- Other

\vspace{0.8em}
\textbf{Q12.} On average, how much time did you previously spend per assignment (assuming a format similar to standard past paper design)? \\
- Less than 5 minutes \\
- 5–10 minutes \\
- 10–30 minutes \\
- More than 30 minutes \\
- Other: \underline{\hspace{2.5cm}}

\vspace{0.8em}
\textbf{Q13.} Compared to your previous method(s), in which aspects does this tool perform better? (Select all that apply)  \\
- Accuracy \\
- Speed \\
- Usability \\
- Reliability \\
- Alignment with curriculum 

\vspace{0.8em}
\textbf{Q14.} If you were to perform similar tasks again, which method would you prefer? \\
- I would continue using this tool \\
- I would return to my previous method(s) \\
- I would combine this tool with previous method(s) \\
- I’m not sure yet

\end{tcolorbox}

\end{document}